\documentclass[10pt,a4paper]{article}

\usepackage{graphicx}
\usepackage{amssymb,amsmath}
\usepackage{a4wide}

\newtheorem{thm}{Theorem}
\newtheorem{defn}{Definition}
\newtheorem{lem}{Lemma}
\newtheorem{remark}{Remark}
\newtheorem{algorithm}{Algorithm}

\begin{document}

\title{Optimal Routing for Decode-and-Forward based Cooperation in Wireless Networks}

\author{Lawrence Ong and Mehul Motani\\
Department of Electrical \& Computer Engineering \\
National University of Singapore\\
Email: \{lawrence.ong, motani\}@nus.edu.sg}
%\and
%\authorblockN{Mehul Motani}
%\authorblockA{Department of Electrical \& Computer Engineering\\
%National University of Singapore\\
%Email: motani@nus.edu.sg}
\date{}

\maketitle

\begin{abstract}
We investigate cooperative wireless relay networks in which the nodes can help each other in data transmission. We study different coding strategies in the single-source single-destination network with many relay nodes.  Given the myriad of ways in which nodes can cooperate, there is a natural routing problem, i.e., determining an ordered set of nodes to relay the data from the source to the destination.  We find that for a given route, the decode-and-forward strategy, which is an information theoretic cooperative coding strategy, achieves rates significantly higher than that achievable by the usual multi-hop coding strategy, which is a point-to-point non-cooperative coding strategy. We construct an algorithm to find an optimal route (in terms of rate maximizing) for the decode-and-forward strategy. Since the algorithm runs in factorial time in the worst case, we propose a heuristic algorithm that runs in polynomial time. The heuristic algorithm outputs an optimal route when the nodes transmit independent codewords.  We implement these coding strategies using practical low density parity check codes to compare the performance of the strategies on different routes.
\end{abstract}


% \begin{center} \bfseries EDICS Cak+2, \dotsc, \mathcal{M}_1.tegory: 3-BBND \end{center}
%
% for peerreview papers, inserts a page break and creates the second title.
% Will be ignored for other modes.
%\IEEEpeerreviewmaketitle

\section{Introduction}
%Wireless communications have been receiving much attention in recent years. 
Research in high data rate wireless systems has enabled applications to go wireless and become more interesting, e.g., wireless Internet access, mobile video conferencing and mobile TV on buses and trains. These applications would have been impossible without high rate wireless transmission links. As many wireless devices are battery operated, power constraint is often imposed on them to make sure that they maintain a certain desired lifespan. In this paper, we investigate optimal routing problem to maximize the transmission rate in the wireless network where there is a power constraint on each node.

%A major difference between the wired and wireless mediums is that the latter is broadcast in nature. 
The wireless channel is inherently broadcast, in that messages sent out by a node are heard by all nodes listening in the same frequency band and in communication range. This opens up opportunities for richer forms of cooperation among the wireless users/nodes. For example, rather than using point-to-point multi-hop routing (a direct adaptation from wired networks), where a node only transmits to the next node in the ``route'', cooperative strategies, such as information theoretic relaying~\cite{covergamal79,xiekumar03,kramergastpar04} and opportunistic routing \cite{biswasmorris04}, could be used. These richer forms of cooperation can lead to efficient distributed algorithms and can increase the end-to-end data rates. The gain from cooperation has been shown in information theoretic analyses~ \cite{ongmotani05a}\cite{ongmotani05b} and demonstrated in practical implementations~\cite{sendonariserkip03,sendonariserkip03b,lee06}.

\begin{figure}[t]
\centering
\includegraphics[width=6cm]{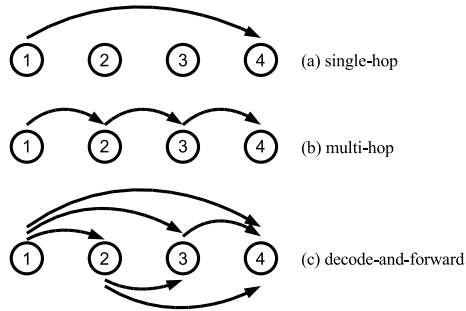}
\caption{Different coding strategies for multiple relay channels.}
\label{fig:different-coding-strategies}
\end{figure}

We now briefly describe what we mean by these richer forms of cooperation, often using the term ``coding'' to highlight that our approach stems from information theory \cite{coverthomas91}.  Figs.~\ref{fig:different-coding-strategies}(a)--(c) depict wireless networks in which node 1 is the source, nodes 2 and 3 are relays, and node 4 is the destination. In Fig.~\ref{fig:different-coding-strategies}(a), since every node can hear what node 1 transmits, the simplest strategy is for node 4 to directly decode from node 1, which we call the \emph{single-hop coding strategy} (SH). However, when nodes 1 and 4 are situated far apart, signals from node 1 go through severe attenuation before they reach node 4. This is when relay nodes 2 and 3 can help. Referring to Fig.~\ref{fig:different-coding-strategies}(b), node 1 transmits to node 2. Node 2 fully decodes the data and re-transmits to node 3. Node 3 does the same and relays the data to node 4. This is the well known \emph{multi-hop coding strategy} (MH). Although we can view relay nodes helping the source to transmit data as a form of cooperation, as far as decoding is concerned, SH and MH are still point-to-point strategies (a node only decodes from one node) and we categorize them as \emph{non-cooperative coding strategies}. Taking a closer look at MH, we see that node 3 can hear and decode node 1's transmission (although it is intended for node 2). This suggests a \emph{cooperative coding strategy}, depicted in Fig.~\ref{fig:different-coding-strategies}(c), in which node 3 decodes transmissions from nodes 1 and 2, and node 4 decodes transmissions from nodes 1--3. This cooperative way of encoding and decoding stems from an information theoretic approach and is termed the \emph{decode-and-forward coding strategy} (DF)~\cite{covergamal79,xiekumar03,kramergastpar04}.

Regardless of whether MH or DF is used for data transmission, there is a sequence of nodes through which data flows. Kurose and Ross~\cite{kuroseross03} define a route as ``the \emph{path} taken by a datagram between source and destination''.  The datagram ``hops'' from one node to the next node, capturing the scenario in which a node receives data only from a node \emph{behind} and forwards data only to the node \emph{in front}. However, in the cooperative coding paradigm, data does not flow from one node to another; rather it is from many to many with complex ways of cooperating. To describe the flow of information in these new modes of cooperation, we define a route as follows.

\begin{defn}
The route taken by a packet from the source to the destination is an ordered set of nodes involved in encoding/transmitting and receiving/decoding of the packet. The sequence of the nodes in the route is determined by the order in which nodes' transmit signals first depend on the packet.
\end{defn}

\begin{remark}
If a group of nodes transmits simultaneously, then they can be ordered arbitrarily within the group. For example, consider a four node network, in which node 1 first broadcasts the message, and then nodes 2 and 3 listen and simultaneously transmit to node 4.  The route here can be described by $\{1,2,3,4 \}$ or $\{1,3,2,4 \}$.
\end{remark}
\begin{remark}
Fig.~\ref{fig:different-coding-strategies} describes three coding strategies for the same four-node network.  The route for SH in Fig.~\ref{fig:different-coding-strategies}a is $\{1,4\}$.  The routes for MH and DF in Figs.~\ref{fig:different-coding-strategies}b \& ~\ref{fig:different-coding-strategies}c respectively are both $\{1,2,3,4\}$.
\end{remark}

Given the myriad of ways in which nodes can cooperate, there is a natural routing problem in the cooperative coding paradigm.  Furthermore, route selection directly affects the end-to-end data transmission rate. For DF, the current routing solutions for MH cannot be applied trivially. In this paper, we construct an algorithm to find an optimal route (in terms of maximizing rates) for DF.

Our contributions in this paper are as follows.
\begin{enumerate}
\item We show how much gain one can expect using DF, a cooperative coding strategy, over MH, a non-cooperative coding strategy, on the same route.
%\item We demonstrate that the average gain in transmission rate for a route using DF over MH increases as the number of the nodes in the route increases.
\item We construct an algorithm that finds rate maximizing routes for DF.
\item We construct a heuristic algorithm that runs in polynomial time. We show that the heuristic algorithm finds an optimal route for DF when the nodes send independent codewords.
\item We implement DF using low-density parity-check (LDPC) codes \cite{gallager62}\cite{mackay99}. Also, we show the performance of codes using different coding strategies and on different routes.
\end{enumerate}

This paper investigates cooperative coding and routing in the wireless network based on an information theoretic approach. A few idealized assumptions are made (e.g., infinite block length, unbounded communication range). Some of these assumptions are, however, relaxed in the simulations in Section~\ref{sec:simulations}.

\subsection{Related Work}

Communications in wireless networks has been progressing from MH to that using cooperative strategies. More research is being directed toward designing codes that are based on information theoretic cooperative coding strategies to harvest the gain in transmission rates predicted by information theory.
Examples of codes based on cooperative coding strategies include DF-based Turbo codes~\cite{zhaovalenti03}\cite{zhangbahceci04} and LDPC codes~\cite{khojastepouraazhang04,chakrabarti06,ezrigastpar06,razaghiyu06} for the single relay channel. It has been mentioned that some of these codes can be extended to the multiple relay channel~\cite{xiekumar03}\cite{kramergastpar04}\cite{ongmotani05a}\cite{ongmotani05b}.

In the past, link optimization (i.e., maximizing the transmission rate between node pairs) and route optimization were done separately.  Routing was optimized after the links between the nodes had been established. Algorithms such as Bellman-Ford \cite[Section 24.1]{cormen01}\cite{suhuanglee04} and Dijkstra's algorithm \cite{dijkstra59} that assign costs to all links were used to find a route with the lowest cost from source to the destination. These ways of separating routing and coding are not optimal for MH or DF as the rates of the links change depending on which route is chosen. Realizing the inter-dependency between links and routes, it has been suggested that links and routes be jointly optimized \cite{cruzsanthanam03,liephremisdes03,lukrishnamachari03,zhangwu05}. This gives rise to cross-layering \cite{srivastavamotani05} in the OSI model. However, in these joint routing and coding work, data transmission from the source to destination is still based on MH. Routing algorithms that are optimized for MH might not be suitable for DF.

In Ad hoc On-demand Distance Vector Routing (AODV)~\cite{perkins97} and Dynamic Source Routing (DSR)~\cite{johnson96}, the source node broadcasts a route discovery packet. Neighboring nodes receive and re-broadcast the packet. When the destination receives the packet, a route is formed by tracing the path that the packet took. These routing algorithms minimize the transmission delay but might not optimize the transmission rate.

In Extremely Opportunistic Routing (ExOR)~\cite{biswasmorris04}, a node broadcasts its data to a set of potential relays.  Nodes in this set transmit acknowledgments and then selected nodes forward the data. Though ExOR does not have predefined routes, MH is used on the \emph{effective} route taken by a packet.

As far as we know, routing algorithms for cooperative coding have not been investigated. In this paper, we propose algorithms to find optimal routes for DF-based codes in the multiple relay channel. Our work complements code design by finding the best route (rate maximizing) on which the codes can be used.
As previous work focused on cooperative coding for the single relay channel, in this paper, we implement DF-based LDPC codes on the multiple relay channel. We then compare the transmission rate of different coding strategies on different routes.

We focus on the multiple relay channel~\cite{xiekumar03}\cite{kramergastpar04}\cite{ongmotani05a}\cite{ongmotani05b}, which is a single-source single-destination network, as a first step towards understanding general multiple-source multiple-destination networks. We study DF because it is one of the ``more implementable'' information theoretic coding strategies~\cite{razaghiyu06,zhaovalenti03,zhangbahceci04,khojastepouraazhang04,chakrabarti06,ezrigastpar06}.

%Our contributions in this paper are as follows.
%\begin{enumerate}
%\item We show how much gain one can expect using DF, a cooperative coding strategy, over MH, a non-cooperative coding strategy, on the same route.
%\item We demonstrate that the average gain in transmission rate for a route using DF over MH increases as the number of the nodes in the route increases.
%\item We construct an algorithm that finds optimal routes (in terms of rate maximizing) for DF.
%\item We construct a heuristic algorithm that runs in polynomial time and finds a route which achieves rate close to  (or as high as) the maximum DF rate.
%\end{enumerate}

\section{Motivating Cooperation} \label{sec:routing_channel_model}
\subsection{Network Model}
We consider a $D$-node network $\mathcal{S} =\{1, 2, 3 \dotsc,$ $D-1, D\}$ with one source (node 1) and one destination (node $D$). 
%The nodes are uniformly distributed over a square area. 
Node $i, \forall i \in \mathcal{S}$, either transmits at fixed average power $P_i$ or turns off. We use the standard path loss model for signal propagation. The received power at node $t$ from node $i$ is given by
%\begin{equation}
$
P_{it} = \kappa d_{it}^{-\eta} P_i,
$
%\end{equation}
where $d_{it}$ is the distance between nodes $i$ and $t$, $\eta$ is the path loss exponent
($\eta \geq 2$ with equality for free space transmission), and $\kappa$ is a positive constant.
The receiver at node $t$ is subject to thermal ambient noise of power $N_t$. 
We assume duplex nodes, i.e., nodes can transmit and receive simultaneously.
We assume that all nodes have the same noise variance.

Given this network model, we investigate how nodes can cooperatively send messages from the source to destination. We study and compare several coding strategies.

\begin{remark}
We consider single-flow networks. This is the first step in understanding a more complicated problem of multiple flows.  The relevance of our work in multiple-flow networks is as follows:
\begin{enumerate}
\item In a multiple-flow networks where each flow uses an allocated orthogonal channel, the rate of each flow can be optimized in respective channel using the algorithm derived in this paper.
\item In a multiple-flow network with existing flows, if we wish to add a new flow, the algorithm in this paper finds an optimal route for the new flow. Note that adding a new flow might affect existing flows. We can restrict the transmit power of nodes in the new flow to control the interference introduced.
\end{enumerate}
\end{remark}

\subsection{Single-Hop Coding Strategy (SH)}
In SH, the source directly transmits data to the destination. The signal-to-noise ratio (SNR) at the destination, node $D$, is given by
%\begin{equation}
$
\gamma_{\text{SH}}(D) = P_{1D}N_D^{-1}.
$
%\end{equation}
The Shannon capacity of this SH link is
%\begin{equation}
$
R_{\text{SH}} = \frac{1}{2} \log \left( 1 + \gamma_{\text{SH}}(D) \right).
$
%\end{equation}
This rate depends on the source-destination distance and can be poor if the source and destination are situated far away from each other (because of signal attenuation).
%This is when the relays come into aid.

\begin{remark}
We assume that nodes that do not participate in relaying the data for a source-destination pair do not cause interference. Another way to account for the external noise is to include it in the receiver noise.
\end{remark}

\subsection{Multi-Hop Coding Strategy (MH)}
In MH, we make use of the relays to aid the transmission from the source to the destination. The source simply transmits to the next relay. The first relay decodes the message and re-transmits it to the second relay, and so on until the destination. This can improve the transmission rate if the attenuation from the source/relay to the next relay is reduced as compared to that in SH. However, since all relays transmit simultaneously, there exists interference, beside noise, at the receiver.

In the rest of this paper, we denote a route by $\mathcal{M} = \{ m_1, m_2, \dotsc, m_{|\mathcal{M}|} \}$.
We define the set of all possible routes from the source (node 1) to the destination (node $D$) by $\Pi (\mathcal{S}) = \Big\{ \{ m_1, m_2, \dotsc, m_{|\mathcal{M}|}\}: m_2,\dotsc,m_{|\mathcal{M}|-1}$ are all possible selections and permutations of the relays (including the empty set), $m_1=1, m_{|\mathcal{M}|}=D \Big\}$.

Using the route $\mathcal{M}$, the SNR at node $m_t$ is
\begin{equation} \label{eq:snr-mh}
\gamma_{\text{MH}}(m_t,\mathcal{M}) = P_{m_{t-1}m_t}\left(\sum\limits_{i=1,i \neq t}^{|\mathcal{M}|-1} P_{m_im_t} + N_{m_t}\right)^{-1}.
\end{equation}
Since all relays and the destination must fully decode the messages, the transmission rate from the source to the destination using route $\mathcal{M}$ is 
\begin{equation}
R_{\text{MH}}(\mathcal{M}) = \min_{ m \in \mathcal{M} \setminus \{m_1\}} \frac{1}{2} \log \left( 1 + \gamma_{\text{MH}}(m,\mathcal{M}) \right).
\end{equation}
We term $R_{\text{MH}}(\mathcal{M})$ the rate \emph{supported} by the route $\mathcal{M}$ using MH.
The maximum rate using by MH, optimized over all possible routes, is
\begin{equation} \label{eq:max-mh-rate}
R_{\text{MH}}^{\text{max}} = \max_{\mathcal{M} \in \Pi(\mathcal{S})} R_{\text{MH}}(\mathcal{M}).
\end{equation}
We note that there may exist more than one route that support this maximum rate.

\subsection{Decode-and-Forward Cooperative Coding (DF)}
Using DF~\cite{covergamal79,xiekumar03,kramergastpar04}, each decoder decodes transmissions from all nodes behind. E.g., the third node in the route decodes the transmissions from the first and the second node. So, the third node decodes each source message using two blocks of received codewords. In addition, assuming that the nodes in front decode the messages correctly, a node knows what they transmit and hence it cancels the interference from these nodes.
It has been shown that in order to maximize the DF rate on route $\mathcal{M}$, node $m_i$ transmits 
%\begin{equation}
$X_{m_i} = \sum_{j=i+1}^{|\mathcal{M}|} \sqrt{\alpha_{m_im_j}P_{m_i}}U_{m_j}$,
%\end{equation}
for $0 \leq \sum_{j=i+1}^{|\mathcal{M}|} \alpha_{m_i m_j} \leq 1$, $\forall i=1, \dotsc, |\mathcal{M}|-1$. $U_{m_j}$ are independent Gaussian random variables with unit variance. $\{\alpha_{ij}|j=i+1, \dotsc, |\mathcal{M}|\}$ are the power splits of node $i$, allocating portions of its transmit power to transmit independent \emph{sub-codewords} $U_j$. Doing this the SNR of node $m_t$ in route $\mathcal{M}$ is
\begin{equation} \label{eq:df-no-coherent}
\gamma_{\text{DF}}( m_t, \mathcal{M}) = N_{m_t}^{-1}\sum_{j=2}^t \left( \sum_{i=1}^{j-1} \sqrt{\alpha_{m_im_j}P_{m_im_t}} \right)^2.
\end{equation}

Using the route $\mathcal{M}$, DF can achieve rates up to
\begin{equation}
R_{\text{DF}}(\mathcal{M}) = \max_{\alpha_{ij}} \min_{ m \in \mathcal{M} \setminus \{ m_1 \}} \frac{1}{2} \log \left( 1 + \gamma_{\text{DF}}(m,\mathcal{M}) \right),
\end{equation}
and the maximum rate using by DF is
\begin{equation} \label{eq:max-df-rate}
R_{\text{DF}}^{\text{max}} = \max_{\mathcal{M} \in \Pi(\mathcal{S})} R_{\text{DF}}(\mathcal{M}).
\end{equation}

\begin{defn}
We define the \emph{reception rate} of node $m$ in route $\mathcal{M}$ as
%\begin{equation}
$
R_m(\mathcal{M}) = \frac{1}{2} \log \left( 1 + \gamma_{\text{DF}}(m,\mathcal{M}) \right).
$
%\end{equation}
It is the rate at which node $m$ can fully decode the messages. The same concept applies to MH.
\end{defn}

\begin{remark}
In practice, a relay in the route might decode a message wrongly, and hence forward the wrong message. When this happens, the nodes behind, when trying to cancel the co-channel interference introduced by this relay, will introduce more noise at their decoders. While this scenario is not captured in \eqref{eq:df-no-coherent}, we allow imperfect interference cancellation in our simulations (Section~\ref{sec:simulations}).
\end{remark}

\begin{figure}[t]
\centering
\includegraphics[width=0.6\linewidth]{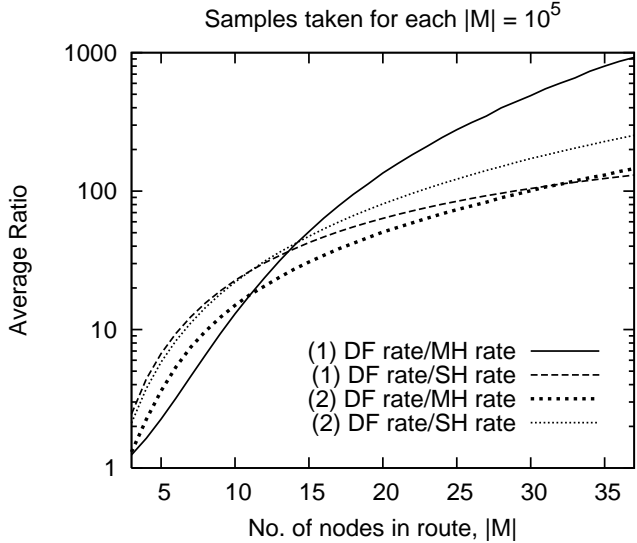}
\caption{Ratio of average transmission rate SH, MH, and DF versus $|\mathcal{M}|$ for two cases: (1) $d_{1D}= 10$ and (2) $d_{1D} = |\mathcal{M}|-1$.}
\label{fig:df-vs-mh}
\end{figure}

\subsection{Comparing the Strategies}

It is easy to see that, for any chosen route $\mathcal{M}$ with four nodes or more,
%\begin{equation}
$
\gamma_{\text{DF}}(m,\mathcal{M}) > \gamma_{\text{MH}}(m,\mathcal{M}), \quad \forall m \in \mathcal{M}.
$
%\end{equation}
Also, we can show that for any $\mathcal{M} \in \Pi(\mathcal{S})$,
\begin{subequations}
\begin{align}
R_{\text{DF}}({\mathcal{M}}) & = R_{\text{MH}}({\mathcal{M}}) = R_{\text{SH}}, & \text{for }|\mathcal{M}|=2,\\
R_{\text{DF}}({\mathcal{M}}) & \geq R_{\text{MH}}({\mathcal{M}}), & \text{for }|\mathcal{M}|=3,\\
R_{\text{DF}}({\mathcal{M}}) & > R_{\text{MH}}({\mathcal{M}}), & \text{for }|\mathcal{M}| \geq 4,
\end{align}
\end{subequations}
and $R_{\text{DF}}^{\text{max}} \geq R_{\text{MH}}^{\text{max}} \geq R_{\text{SH}}$.

However, it is not clear how much again, on average, we can expect using DF compared to MH and SH. Now, we compare the rates of SH, MH, and DF for randomly generated routes of different lengths in a line topology. We consider two cases: (1) $d_{1D}= 10$ and (2) $d_{1D} = |\mathcal{M}|-1$. Then $|\mathcal{M}|-2$ nodes are randomly placed along the straight line joining nodes 1 and $D$.  Note that in case 1, node density increases with the number of nodes while in case 2, the average adjacent node spacing is constant for all $|\mathcal{M}|$.
We set $P_i = N_i = 1$ for all transmitters and receivers, and $\kappa=1, \eta=2$.
For each randomly generated route, we calculate the transmission rate using SH, MH, and DF. Here we restrict the nodes to transmit independent codewords for easier optimization, i.e., we set $\alpha_{ij}=1, \forall i, \forall j=i+1$ and $\alpha_{ij}=0, \forall j \neq i+1$. 

Fig~\ref{fig:df-vs-mh} shows the results for cases 1 and 2. For a route of 25 nodes, the DF rate is roughly two orders of magnitude higher than that of SH and MH for both cases. Moreover, as we increase the number of nodes in the route, the gain of DF over MH/SH increases for both cases.

We note that if the nodes are allowed to send arbitrarily correlated codewords, the DF rate can be higher. For example, consider the route $\mathcal{M} = \{ (0,0), (0.5,0), (2,0), (3,0), (4,0) \}$. With independent codewords,\\ $R_{\text{DF}}(\mathcal{M}) /R_{\text{MH}}(\mathcal{M}) = 2.95$, but with arbitrarily correlated codewords,\\ $R_{\text{DF}}(\mathcal{M}) /R_{\text{MH}}(\mathcal{M}) = 4.40$.
%MH = 0.130031
%DFic = 0.380406
%DF = 0.5723000

\section{The Optimal Routing Problem}
We define the optimal route set for DF as
\begin{equation}
\mathcal{Q}_{\text{DF}}  \triangleq \{\mathcal{M}\in\Pi(\mathcal{S}) : R_{\text{DF}}(\mathcal{M})=R_{\text{DF}}^{\text{max}}\}\,
\end{equation}
where $\Pi (\mathcal{S})$ is a set of all possible routes from the source to the destination.
We define the optimal route set because the rate maximizing route may not be unique. Then the optimal DF routing problem is
\begin{equation}
\text{Find at least one~} \mathcal{M}_{\text{DF}}^{\text{opt}} \in \mathcal{Q}_{\text{DF}}
\text{~and~} R_{\text{DF}}\left( \mathcal{M}_{\text{DF}}^{\text{opt}} \right). \nonumber
\end{equation}
The optimal route set and routing problem for MH are similarly defined.

Finding optimal routes for MH and DF by brute force is hard, as it involves testing all routes in $\Pi (\mathcal{S})$.
In Sections~\ref{sec:routing_nna} and \ref{sec:routing_nnsa}, we construct an algorithm that finds $\mathcal{M}_{\text{DF}}^{\text{opt}}$, potentially without having to test all routes in $\Pi(\mathcal{S})$. However, this algorithm runs in factorial time in the worst case. Hence, in Section~\ref{sec:NWCA},  we proposed a heuristic algorithm, which runs in polynomial time.
%\begin{equation}
%\text{Find some~} \mathcal{M}' \in \Pi(\mathcal{S})
%\text{~and~} R_{\text{DF}}\left( \mathcal{M}' \right) \nonumber
%\end{equation}
%in polynomial time and
%\begin{equation}
%\text{Compare~} R_{\text{DF}}\left( \mathcal{M}_{\text{HEU}} \right)
%\text{~with~} R_{\text{DF}}^{\text{max}}. \nonumber
%\end{equation}
%In Section~\ref{sec:NWCA}, we propose a heuristic algorithm, which finds a route in polynomial time that achieves rate close to the optimal rate $R_{\text{DF}}^{\text{max}}$.

\section{The Nearest Neighbor Algorithm}\label{sec:routing_nna}
Now, we present an algorithm to find an optimal route for DF. In the section, we assume that nodes use independent codewords, i.e., we set $\alpha_{ij}=1, \forall i, \forall j=i+1$ and $\alpha_{ij}=0, \forall j \neq i+1$. 

\begin{remark}
Although the theorems in this section are proven assuming that the nodes send independent codewords, we can also show that they hold even when the nodes transmit arbitrarily correlated codewords~\cite{ongmotani07isit}.
\end{remark}

\begin{remark}
We consider independent codewords in this section because \emph{coherent combining} is practically infeasible. When the nodes are operating in the GHz range, it is difficult, if not impossible, to synchronize the carriers to nanosecond accuracy. Furthermore, even if we have very precise clocks, coherent combining is still unlikely in a multiple-node network. For example, in a four-node route, even if we manage to synchronize nodes 1 and 2 to allow coherent combining at node 3, we might not be able to ensure that they will coherently combine at node 4.
\end{remark}

First, we define the \emph{nearest neighbor} with respect to a route.
\begin{defn}
Node $i \notin \mathcal{M}$ is a nearest neighbor with respect to the route $\mathcal{M}$ iff
\begin{equation}\label{eq:nearest_neighbor_condition}
P_{mi} \geq P_{mj}, \quad \forall m \in \mathcal{M}, \forall j \in \mathcal{S} \setminus (\mathcal{M} \cup \{ i \}).
\end{equation}
\end{defn}
\smallskip
\noindent Note that nearest neighbor might not be unique. 
Now, we describe the nearest neighbor algorithm (NNA).
\begin{algorithm}[NNA]
\mbox{\,}
\begin{enumerate}
\item  Initialize $\mathcal{M} = \{ m_1\}$, where $m_1=1$.
\item \label{item:pick_neighbor} 
%to the current route $\mathcal{M}$, i.e, we pick the unique nearest neighbor $i \notin \mathcal{M}$ such that
%\begin{equation}
%d_{mi} \leq d_{mj}, \quad \forall m \in \mathcal{M}, \forall j \in \mathcal{S} \setminus ( \mathcal{M} \cup \{ i \}),
%\end{equation}
%with at least one strict inequality.
%We assume that such a unique nearest neighbor exists. 
If there exists a unique nearest neighbor $i^*$ with respect to the current route $\mathcal{M}$, we append $i^*$ to the current route: 
$\mathcal{M} \leftarrow \mathcal{M} \cup \{ i^* \}$. Else, the NNA terminates prematurely. Since $\mathcal{M}$ is an ordered set, the notation $\mathcal{A} \cup \mathcal{B}$ means appending ordered set $\mathcal{B}$ to the end of ordered set $\mathcal{A}$.
\item Step~\ref{item:pick_neighbor} is repeated until the destination node, node $D$, is added into $\mathcal{M}$.
\end{enumerate}
\end{algorithm}

The algorithm is said to terminate normally if node $D$ is added to the route. Otherwise, the algorithm is said to terminate prematurely. If the NNA terminates normally, we have the following theorem.

\begin{thm} \label{thm:NNA}
Consider a multiple node wireless network with one source and one destination. If the NNA terminates normally, then the NNA route is optimal for DF.
\end{thm}

To prove Theorem~\ref{thm:NNA}, we need the following lemmas.

\begin{lem}\label{lem:nearest_neighbor}
When we add the unique nearest neighbor, node $a^*$, to route $\mathcal{M}$, the rate supported by the new route $\mathcal{M}_1 = \mathcal{M} \cup \{m_{|{M}_1|}=a^*\}$ is greater or equal than the rate supported by the route formed by adding any other node to $\mathcal{M}$.
%$\mathcal{M}_2 = \mathcal{M} \cup \{m_{|{M}_2|}=b\}$. 
%We use superscript $^*$ to indicate a nearest neighbor. 
Mathematically,
\begin{equation}
R_{\text{DF}}(\mathcal{M} \cup \{a^*\}) \geq R_{\text{DF}}(\mathcal{M} \cup \{b\}), \forall b \in \mathcal{S} \setminus ( \mathcal{M} \cup \{a^*\} ).
\end{equation}
\end{lem}

\emph{Proof:} [Proof for Lemma~\ref{lem:nearest_neighbor}] Considering $\mathcal{M}_2$, the reception rate of node $m_{|{M}_2|}=b$ is
\begin{equation}
R_b(\mathcal{M} \cup \{b\}) = \frac{1}{2} \log \left[ 1 + {N_b}^{-1}\sum_{i=1}^{|\mathcal{M}|} P_{m_ib}\right],
\end{equation}
and the reception rate of the node $m_{|{M}_1|}=a^*$ in route $\mathcal{M}_1$ is
\begin{equation}
R_{a^*}(\mathcal{M} \cup \{a^*\}) = \frac{1}{2} \log \left[ 1 + N_{a^*}^{-1}\sum_{i=1}^{|\mathcal{M}|}P_{m_ia^*}\right].
\end{equation}
Clearly, if $P_{m a^*} \geq P_{m b}, \quad \forall m \in \mathcal{M}$ with at least one inequality, $R_{a^*}(\mathcal{M}_1) > R_b(\mathcal{M}_2)$. Hence $R_{\text{DF}}(\mathcal{M}_1) \geq R_{\text{DF}} (\mathcal{M}_2)$. Hence we have Lemma~\ref{lem:nearest_neighbor}.

We have proven that at any point of time during route construction, in order to maximize the rate supported by the route, we must choose the nearest neighbor (assuming it exists). Next, we show that choosing the nearest neighbor will not harm the rate supported by the route even when more nodes are added.

\begin{lem}\label{lem:nearest_neighbor_in_middle}
Let $\mathcal{M} = \{ a_1^*, a_2^*, \dotsc, a_{|\mathcal{M}|}^*\}$
be a route formed by adding the nearest neighbor one by one starting from the source. Now, arbitrarily add $K$ nodes to $\mathcal{M}$. The first node $b_1$ is not a nearest neighbor and the rest may or may not be nearest neighbors. In other words,
%\begin{equation}
$
\mathcal{M}_1 = \{ a_1^*, a_2^*, \dotsc, a_{|\mathcal{M}|}^*, b_1, b_2, \dotsc, b_K \},
$
%\end{equation}
where $b_1$ is not a nearest neighbor to $\mathcal{M}$. We can always replace $b_1$ by the nearest neighbor $a_{{|\mathcal{M}|}+1}^*$ (assuming it exists) to obtain
\begin{equation} \label{eq:lemma-2-eq}  \scriptstyle
\mathcal{M}_2 = 
\begin{cases}
\{ a_1^*, \dotsc, a_{|\mathcal{M}|}^*, a_{{|\mathcal{M}|}+1}^*, b_1, \dotsc, b_{K-1} \},\\
\quad\quad \text{if } a_{{|\mathcal{M}|}+1}^* \notin \{b_1, \dotsc, b_{K-1} \}, \\
\{ a_1^*, \dotsc, a_{|\mathcal{M}|}^*, a_{{|\mathcal{M}|}+1}^*, b_1, \dotsc, b_{k-1}, b_{k+1}, \dotsc, b_K \},\\
\quad\quad \text{if $a_{{|\mathcal{M}|}+1}^* = b_k$, for some $b_k \in \{b_1, \dotsc, b_{K-1} \},$}
\end{cases}
\end{equation}
where
%\begin{equation}
$
R_{\text{DF}}(\mathcal{M}_2) \geq R_{\text{DF}}(\mathcal{M}_1).
$
%\end{equation}
\end{lem}

\emph{Proof:} [Proof for Lemma~\ref{lem:nearest_neighbor_in_middle}] For both cases in \eqref{eq:lemma-2-eq}, the reception rates for the first $|\mathcal{M}|$ nodes in both $\mathcal{M}_1$ and $\mathcal{M}_2$ remain the same as each of them decodes from the same nodes behind the route. In equations,
\begin{equation}
R_{a_i^*} (\mathcal{M}_2) =  R_{a_i^*} (\mathcal{M}_1), \quad \forall i = 2, 3, \dotsc , |\mathcal{M}|.
\end{equation}
From Lemma~\ref{lem:nearest_neighbor}, $R_{a_{{|\mathcal{M}|}+1}^*} (\mathcal{M}_2) > R_{b_1} (\mathcal{M}_1)$.

Now, we study the case when $a_{{|\mathcal{M}|}+1}^* \notin \{b_1, \dotsc, b_{K-1} \}$. For nodes $\{ b_1, \dots, b_{K-1}\}$ in $\mathcal{M}_2$, with an additional node behind, i.e., $a_{{|\mathcal{M}|}+1}^*$, the reception rates of these nodes are higher than the same nodes in $\mathcal{M}_1$:
\begin{equation} \label{eq:explain4}
R_{b_i} (\mathcal{M}_2)  > R_{b_i} (\mathcal{M}_1), \quad \forall i = 1, 2, \dotsc , K-1. 
\end{equation}

Now, we study the case when $a_{{|\mathcal{M}|}+1}^* = b_k$ for some $b_k \in \{b_1, \dotsc, b_{K-1} \}$. Similar to the first case, with an additional transmitting node $a_{{|\mathcal{M}|}+1}^*$, the nodes $\{ b_1, \dots, b_{k-1}\}$ in $\mathcal{M}_2$ have higher reception rates compared to those in $\mathcal{M}_1$, i.e.,
\begin{equation}
R_{b_i} (\mathcal{M}_2)  > R_{b_i} (\mathcal{M}_1), \quad \forall i = 1, 2, \dotsc , k-1. 
\end{equation}
For nodes $\{ b_{k+1}, \dotsc, b_K \}$, there is no change in the reception rate because each of them has exactly the same nodes behind them in both $\mathcal{M}_2$ and $\mathcal{M}_1$. So,
\begin{equation}
R_{b_i} (\mathcal{M}_2) = R_{b_i} (\mathcal{M}_1), \quad \forall i = k+1, k+2, \dotsc , K.
\end{equation}

Lemma~\ref{lem:nearest_neighbor_in_middle} follows by noting that
$R_{\text{DF}}(\mathcal{M}_2) = \min_{i \in \mathcal{M}_2 \setminus \{ a^*_1\}} R_i(\mathcal{M}_2)
\geq \min_{i \in \mathcal{M}_1 \setminus \{ a^*_1\}} R_i(\mathcal{M}_1)
= R_{\text{DF}}(\mathcal{M}_1).$
%\begin{subequations}
%\begin{align}
%R_{\text{DF}}(\mathcal{M}_2) & = \min_{i \in \mathcal{M}_2 \setminus \{ a^*_1\}} R_i(\mathcal{M}_2) \\
%& \geq \min_{i \in \mathcal{M}_1 \setminus \{ a^*_1\}} R_i(\mathcal{M}_1) \\
%& = R_{\text{DF}}(\mathcal{M}_1).
%\end{align}
%\end{subequations}
%This proves Lemma~\ref{lem:nearest_neighbor_in_middle}.

\begin{lem}\label{lem::same_length_nna_best}
For a route that contains all nearest neighbors, the supported rate is always higher or equal to any route, of the same length, with one or more non-nearest neighbors in it.
\end{lem}

\emph{Proof:} [Proof for Lemma~\ref{lem::same_length_nna_best}]
Lemma~\ref{lem::same_length_nna_best} can be proven by applying Lemma~\ref{lem:nearest_neighbor_in_middle} recursively until the entire set is replaced by nearest neighbor nodes. 
%Lemma~\ref{lem::same_length_nna_best} can be proven by applying Lemma~\ref{lem:nearest_neighbor_in_middle} recursively. Starting from the second node onwards, we replace the first non-nearest neighbor with a nearest neighbor. We remove the last node if the resulting set if longer. In each step, the supported rate can only increase. We do that until the entire set is replaced by nearest neighbor nodes. %Hence, we get Lemma~\ref{lem::same_length_nna_best}.

Now we consider routes of different lengths but that end on the same node.
\begin{lem} \label{lem:NNA_arbitrary_length}
Consider a route $\mathcal{M}_1$, where node 1 is the source, and node $m_{|{M}_1|} = D$ is the destination, that is
\begin{equation}
\mathcal{M}_1 = \{ m_1^*, m_2, \dotsc, m_{|{M}_1|} \}.
\end{equation}
Here, one or more nodes in $\{m_2, \dotsc, m_{|{M}_1|}\}$ are not nearest neighbors.
The following route where all nodes are added according to the NNA (assuming that it does not terminate prematurely) supports rate as good or higher than that supported by $\mathcal{M}_1$.
\begin{equation}
\mathcal{M}_2 = \{ m_1^*, m_2^*, \dotsc, m_{|{M}_2|}^*\},
\end{equation}
where $m_{|{M}_2|}^* = D$ and ${|{M}_1|}$ not necessarily equals ${|{M}_2|}$. In other words,
%\begin{equation}
$
R_{\text{DF}}(\mathcal{M}_2) \geq R_{\text{DF}}(\mathcal{M}_1).
$
%\end{equation}
\end{lem}

\emph{Proof:} [Proof for Lemma~\ref{lem:NNA_arbitrary_length}] First of all, we consider the case ${|{M}_1|}={|{M}_2|}$. The results follows immediately from Lemma~\ref{lem::same_length_nna_best}. Second, we consider ${|{M}_1|} > {|{M}_2|}$. We consider first ${|{M}_2|}$ nodes in $\mathcal{M}_1$, i.e., 
$\mathcal{M}_1' = \{ m_1^*, m_2, \dotsc, m_{|{M}_2|}\}$.
Then,
\begin{equation}
R_{\text{DF}}(\mathcal{M}_2) \geq R_{\text{DF}}(\mathcal{M}_1') \geq R_{\text{DF}}(\mathcal{M}_1).
\end{equation}
The first inequality is obtained by applying Lemma~\ref{lem::same_length_nna_best}. $\mathcal{M}_2$ and $\mathcal{M}_1'$ are of the same length. The former is formed using the NNA while the latter is not. The second inequality can be argued as follows. The first $|\mathcal{M}_2|$ nodes in both routes $\mathcal{M}_1'$ and $\mathcal{M}_1$ are identical. Hence the reception rates are the same. However, there are additional nodes in $\mathcal{M}_1$ whose reception rate might be lower than $R_{\text{DF}}(\mathcal{M}_1')$. Hence, the rate supported by $\mathcal{M}_1'$ can only be higher than that of $\mathcal{M}_1$.

Lastly, consider ${|{M}_2|} > {|{M}_1|}$. We replace the transmitting nodes in $\mathcal{M}_1$ with nearest neighbors and obtain
\begin{equation}
\mathcal{M}_3 = \{ m_1^*, m_2^*, \dotsc, m_{|{M}_1|-1}^*, m_{|{M}_1|}=D \}.
\end{equation}
Note that $m_{|{M}_1|}$ might not be the nearest neighbor. Clearly, using Lemma~\ref{lem:nearest_neighbor_in_middle},
$R_{\text{DF}}(\mathcal{M}_3) \geq R_{\text{DF}}(\mathcal{M}_1)$.
Now,
\begin{subequations} 
\begin{align}  \scriptstyle
& R_{\text{DF}}(\mathcal{M}_2)\nonumber\\
& = \min \{ R_{m_2^*}(\mathcal{M}_2), \dotsc, R_{m_{|{M}_1|-1}^*}(\mathcal{M}_2), R_{m_{|{M}_1|}^*}(\mathcal{M}_2), \nonumber\\
& \quad \dotsc, R_{m_{|{M}_2|-1}^*}(\mathcal{M}_2),  R_{m_{|{M}_2|}^*}(\mathcal{M}_2) =R_D(\mathcal{M}_2) \},\\
& \geq \min \{ R_{m_2^*}(\mathcal{M}_3), \dotsc, R_{m_{{|{M}_1|}-1}^*}(\mathcal{M}_3), R_D(\mathcal{M}_3) \}, \label{lemma4-1}\\
& = R_{\text{DF}}(\mathcal{M}_3) \geq R_{\text{DF}}(\mathcal{M}_1).
\end{align}
\end{subequations}

The inequality in \eqref{lemma4-1} is because in $\mathcal{M}_2$, $\{ m^*_{|\mathcal{M}_1|}, \dotsc, m^*_{|\mathcal{M}_2|-1} \}$ are added to $\{m_1^*, \dotsc, m^*_{|\mathcal{M}_1|-1}\}$ before $D$. A necessary condition for this is
\begin{multline}
P_{m n} \geq P_{m D}, \quad \forall m \in  \{m_1^*, \dotsc, m^*_{|\mathcal{M}_1|-1}\},\\ \forall n \in \{ m^*_{|\mathcal{M}_1|}, \dotsc, m^*_{|\mathcal{M}_2|-1} \},
\end{multline}
with at least one strict inequality for each $n$.
Hence, $R_n(\mathcal{M}_2) > R_D(\mathcal{M}_3), \forall n \in \{ m^*_{|\mathcal{M}_1|}, \dotsc, m^*_{|\mathcal{M}_2|-1} \}$.
With additional nodes transmitting to $D$ in $\mathcal{M}_2$,
$R_D(\mathcal{M}_2) > R_D(\mathcal{M}_3)$.
Hence, we have Lemma~\ref{lem:NNA_arbitrary_length}.

\emph{Proof:} [Proof for Theorem~\ref{thm:NNA}]
From Lemma~\ref{lem:NNA_arbitrary_length}, we know that if the NNA terminates normally, the route (from the source to the destination) formed using the NNA can support transmission rates as high as any other route. In other words, the NNA finds a route that supports the highest rate achievable by DF. Theorem~\ref{thm:NNA} follows.

\begin{remark}
We note the NNA terminates normally if and only if a unique nearest neighbor exists at each step. In the next section, we extend the NNA to an algorithm which terminates normally given any network topology.
\end{remark}

\section{The Nearest Neighbor Set Algorithm}\label{sec:routing_nnsa}
In this section, we modify the NNA so that it terminates normally in any multiple node wireless network with a single source and a single destination. We term this algorithm the nearest neighbor set algorithm (NNSA). First, we define the \emph{nearest neighbor set}.
\begin{defn}
The nearest neighbor set $\mathcal{N} = \{ n_1, n_2, \dotsc, n_{|{N}|} \}$ with respect to route $\mathcal{M} = \{ m_1, m_2,$ $\dotsc, m_{|\mathcal{M}|} \}$ is defined as the smallest set $\mathcal{N}$ where each $n \in \mathcal{N} \subseteq \mathcal{S} \setminus \mathcal{M}$ satisfies the following condition.
\begin{equation}
P_{mn} \geq P_{ma}, \quad \forall m \in \mathcal{M}, \forall a \in \mathcal{S} \setminus( \mathcal{M} \cup \mathcal{N}),
\end{equation}
with at least one strict inequality for every pair of $(n,a) \in \{ (n,a) : n \in \mathcal{N}, a \in \mathcal{S} \setminus (\mathcal{M} \cup \mathcal{N}) \}$.
\end{defn}

Now we describe the NNSA.
%\begin{enumerate}
%\item Starting with the source node, we have $\mathcal{M} = \{ 1 \}$.
%\item \label{item:NNSA_add_nodes}
%\begin{enumerate}
%\item If a unique nearest neighbor $i^*$ exists, we add it to the route:
%\begin{equation}
%\mathcal{M} \leftarrow \mathcal{M} \cup \{ i^* \}.\label{eq:NNSAa}
%\end{equation}
%\item Else, we find the nearest neighbor set $\mathcal{N}$. Now, each element in $\mathcal{N}$ is added to the end of $\mathcal{M}$ to form one new route. The original route $\mathcal{M}$ branches out to $|\mathcal{N}|$ routes as follows:
%\begin{equation}
%\mathcal{M}_i \leftarrow \mathcal{M} \cup \{ n_i \}, \quad i=1, \dotsc, |\mathcal{N}|. %\label{eq:NNSA}
%\end{equation}
%\end{enumerate}
%\item For the new route in \eqref{eq:NNSAa} or for each new route in \eqref{eq:NNSA}, step~\ref{item:NNSA_add_nodes} is repeated until the destination is added to all routes.
%\end{enumerate}

\begin{algorithm}[NNSA]
\mbox{\,}
\begin{enumerate}
\item Starting with the source node, we have $\mathcal{M} = \{ 1 \}$.
\item \label{item:NNSA_add_nodes}
Find the nearest neighbor set $\mathcal{N}$. 
%Now, each element in $\mathcal{N}$ is added to the end of $\mathcal{M}$ to form one new route. 
The original route $\mathcal{M}$ branches out to $|\mathcal{N}|$ new routes as follows:
\begin{equation}
\mathcal{M}_i \leftarrow \mathcal{M} \cup \{ n_i \}, \quad i=1, \dotsc, |\mathcal{N}|. \label{eq:NNSA}
\end{equation}
\item For each new route in \eqref{eq:NNSA}, step~\ref{item:NNSA_add_nodes} is repeated until the destination is added to all routes.
\end{enumerate}
\end{algorithm}

When the algorithm terminates, we end up with many routes from the source to the destination. We term these routes \emph{NNSA candidates}. We calculate the supported rate of each candidate and choose the one which gives the highest supported rate. The following theorem says that any NNSA candidate that gives the highest supported rate is an optimal route for DF.

\begin{thm} \label{thm:NNSA}
Consider a single-source single-destination multiple node wireless network. The NNSA candidate routes that give the highest supported rate are optimal for DF.
\end{thm}

\emph{Proof:} [Sketch of proof for Theorem~\ref{thm:NNSA}] 
%The proof is similar to that of Theorem~\ref{thm:NNA} and hence will be omitted. 
Using the technique used in the proof of Theorem~\ref{thm:NNA}, we can show that adding a node that does not belong to the nearest neighbor set can only be suboptimal. We can always replace that node with one from the nearest neighbor set and obtain an equal or higher rate. In other words, we can show that the supported rate of
$\mathcal{M}_1 = \{ 1, m_2, \dotsc, m_{|{M}_1|-1}, D \}$,
with one or more nodes in $\{m_2, \dotsc, m_{|{M}_1|}\}$ not from the nearest neighbor set, is lower or equal to the supported rate of
$\mathcal{M}_2 = \{ 1, m_2^*, \dotsc, m_{|{M}_2|-1}^*, D\}$,
where all nodes in $\mathcal{M}_2$ are added according to the NNSA. In other words,
$R_{\text{DF}}(\mathcal{M}_2) \geq R_{\text{DF}}(\mathcal{M}_1)$.

The NNSA finds all possible routes for which every node is added from the nearest neighbor set. Hence one or more of the NNSA candidates must achieve the highest rate achievable by DF.
This gives us Theorem~\ref{thm:NNSA}.

\begin{remark}\label{ref:sor}
We can show that a shortest optimal route, defined as some $\mathcal{M}_{\text{DF}}^{\text{SOR}} \in \mathcal{Q}_{\text{DF}}$, s.t. $\lvert \mathcal{M}_{\text{DF}}^{\text{SOR}} \rvert \leq |\mathcal{M}|, \forall \mathcal{M} \in \mathcal{Q}_{\text{DF}}$, is contained in one of the NNSA candidates that supports $R_{\text{DF}}^{\text{max}}$.
\end{remark}

\begin{remark}
The NNSA might output optimal routes that include more nodes from the network unnecessarily. In other words, shorter optimal routes exist. However, from Remark~\ref{ref:sor}, we can find the shortest optimal route by pruning the optimal routes output by the NNSA.
\end{remark}

\section{Complexity of NNSA} \label{sec:routing_simulation}
With the NNSA, we can now search for the optimal route in the NNSA candidate set, as compared to searching in $\Pi(\mathcal{S})$ using brute force. The number of candidates determines the number of routes whose rate we need to calculate to find optimal routes for DF. We note that the size of the NNSA candidate set might still, in the worst case, equal $|\Pi(\mathcal{S})|$. Using brute force, the number of permutations we need to check is
%\begin{equation}
$|\Pi(\mathcal{S})| = 
1 + {\binom{D-2}{1} + \binom{D-2}{2} + \dotsm \binom{D-2}{D-2}}
= O((D-1)!),
%\end{equation}
$
%\begin{subequations}
%\begin{align}
%|\Pi(\mathcal{S})| & =  1 + \binom{D-2}{1} + \binom{D-2}{2} + \dotsm \binom{D-2}{D-2} \\
%& = O((D-1)!),
%\end{align}
%\end{subequations}
where $D$ is the total number of nodes in the network and
$\binom{n}{k} = \frac{n \times (n-1) \times \dotsm \times 1}{(n-k) \times (n-k-1) \times \dotsm \times 1}.$

We ran the NNSA on 10000 randomly generated networks with a varying number of nodes uniformly distributed in a 1m$\times$1m square area. The source, relays and destination were randomly assigned. On average, half of the NNSA candidate set sizes were less than 0.715\% of the size of $\Pi(\mathcal{S})$ for the 8-node channel and less than 0.253\% for the 11-node channel.

We note that the average size of the NNSA candidate set does grow factorially with the number of nodes. However this does increase the range of finite size networks for which we can find optimal routes.
Furthermore, the NNSA provides insights for designing heuristic algorithms to find good routes for DF based codes. In the next section, we propose heuristic algorithms which find routes in polynomial time.

\begin{remark}
The NNSA builds routes from the source. Relays are added to the route regardless of where the destination is. We will use this observation in designing heuristic algorithms.
\end{remark}

%\section{The Nearest Weighted-Center Algorithm} \label{sec:NWCA}
\section{Heuristic Algorithms} \label{sec:NWCA}
In the NNSA, the optimal route is constructed by adding the ``next hop'' node one by one to the \emph{partial} route. The node to be added is from the nearest neighbor set. If the nearest neighbor set contains more than one node, the current route branches to more than one route, leading to a possibly large NNSA candidate set size. 

To avoid this, we consider a heuristic approach that starts from the source node and repeatedly adds only one ``good'' candidate from the nearest neighbor set until the destination is reached. For the choice of the next hop node, we consider the node which receives the largest sum of received power from all the node  in the existing route. We call this the maximum sum-of-received-power algorithm (MSPA).  By choosing only one node to be added to the partial route, we prevent the algorithm from branching out to multiple routes. This heuristic approach yields only one route, regardless of the network size.  We now explicitly describe the MSPA. 

\begin{algorithm}[MSPA]
\mbox{\,}
\begin{enumerate}
\item Start with the source node: $\mathcal{M}= \{1\}$.
\item \label{item:mspa1} For every node $t \in \mathcal{S} \setminus \mathcal{M}$, find the sum of received power from all nodes in $\mathcal{M}$ to $t$, $\sum_{i \in \mathcal{M}} P_{it}$.
\item \label{item:mspa2} Let $a^*$ be {\em any} node with the highest sum of received power, i.e., $\sum_{i \in \mathcal{M}} P_{ia^*} \geq \sum_{j \in \mathcal{M}} P_{jt}, \forall t \in \mathcal{S} \setminus \mathcal{M}$.
Append node $a^*$ to the route: 
$\mathcal{M} \leftarrow \mathcal{M} \cup \{ a^* \}$. 
%If there are two nodes with the highest sum-of-received-power, randomly choose one.
\item Repeat steps \ref{item:mspa1}--\ref{item:mspa2} until the destination is added to the route.
\end{enumerate}
\end{algorithm}

\begin{remark}
Assuming that the value of the previous sum-of-received-power computations are cached, the complexity of step~\ref{item:mspa1} in MSPA is $O(D)$ because there are at most $(D-1)$ nodes not in the route.  The complexity of the comparisons in step \ref{item:mspa2} is $O(D)$.  Steps \ref{item:mspa1}--\ref{item:mspa2} are repeated at most $(D-1)$ times, giving a worst case complexity of the MSPA of $O(D^2)$. Recall that $D = |\mathcal{S}|$.
\end{remark}

It turns out that the MSPA is optimal if the nodes are restricted to sending independent codewords, as proven in the following theorem.
\begin{thm}\label{thm:mspa}
In a single-source single-destination multiple node wireless network in which the nodes send independent codewords, the MSPA route is optimal for DF.
\end{thm}

\emph{Proof:} [Proof for Theorem~\ref{thm:mspa}]
Consider an optimal route\\ $\mathcal{M}_1 = \{ m_1^*, m_2^*, \dotsc, m_k^*, m_{k+1}^*, \dotsc,  m_{|\mathcal{M}|}^* \}$. Suppose that the first $k$ nodes of the MSPA route are the same as this optimal route but the $(k+1)$-th node is different, i.e., the MSPA route is $\mathcal{M}_2 = \{ m_1^*, m_2^*, \dotsc, m_k^*, a, \dotsc \}$ where $a \neq m_{k+1}^*$.

Since $a$ is added to the route by MSPA, a necessary condition is\\ $\sum_{i=1}^{k} P_{m_i^*a} \geq \sum_{j=1}^k P_{m_j^*m_{k+1}^*}$. So,
\begin{equation}
R_a(\mathcal{M}_2) \geq R_{m_{k+1}^*}(\mathcal{M}_1).
\end{equation}

Now, consider the case where $a \neq m_i^*, \forall i=k+2, \dotsc, |\mathcal{M}|$. We add $a$ to $\mathcal{M}_1$ and obtain $\mathcal{M}_3 = \{ m_1^*, m_2^*, \dotsc, m_k^*, a,  m_{k+1}^*, \dotsc,  m_{|\mathcal{M}|}^* \}$. Then,
\begin{subequations}
\begin{align}
R_{m_i^*}(\mathcal{M}_3)  & = R_{m_i^*}(\mathcal{M}_1), \quad i = 2, \dotsc, k\\
R_a(\mathcal{M}_3) & \geq R_{m_{k+1}^*}(\mathcal{M}_1)\\
R_{m_i^*}(\mathcal{M}_3) & > R_{m_i^*}(\mathcal{M}_1), \quad i = k+1, \dotsc, |\mathcal{M}|.
\end{align}
\end{subequations}
So, $R_\text{DF}(\mathcal{M}_3) \geq R_\text{DF}(\mathcal{M}_1)$.

Suppose $a = m_n^*$, for some $n \in \{k+2, \dotsc, |\mathcal{M}| \}$. We swap the position of $a$ and obtain $\mathcal{M}_4 = \{ m_1^*, m_2^*, \dotsc, m_k^*, a,  m_{k+1}^*, \dotsc, m_{n-1}^*, m_{n+1}^*,  m_{|\mathcal{M}|}^* \}$. Then,
\begin{subequations}
\begin{align}
R_{m_i^*}(\mathcal{M}_4)  & = R_{m_i^*}(\mathcal{M}_1), \quad i = 2, \dotsc, k, n+1, \dotsc, |\mathcal{M}|\\
R_a(\mathcal{M}_4) & \geq R_{m_{k+1}^*}(\mathcal{M}_1)\\
R_{m_i^*}(\mathcal{M}_4) & > R_{m_i^*}(\mathcal{M}_1), \quad i = k+1, \dotsc, n-1.
\end{align}
\end{subequations}
So, $R_\text{DF}(\mathcal{M}_4) \geq R_\text{DF}(\mathcal{M}_1)$.

In summary, we choose an optimal route. Starting from the second node, we compare the optimal route with the MSPA route. If the nodes are different, we insert (or swap, if the node is in the optimal route but at a different position) the node in the MSPA route to the optimal route to obtain a new optimal route. Repeat this by comparing the new optimal route to the MSPA route and changing the first node that differ until the MSPA is contained in an optimal route. Then, we have Theorem~\ref{thm:mspa}.

\begin{remark}
However, unlike the NNA and the NNSA, the MSPA does not output an optimal route when the nodes are allowed to send arbitrarily correlated codewords.  Consider a four-node network with node coordinates 1(0,0), 2(0.418,0), 3(0.209,0.6755), and 4(0.995,0). Assume $P_i=1, N_i=1, \kappa =1,\eta =2$. The MSPA route is $\mathcal{M}_1 = \{1,2,4\}$. The NNSA outputs $\mathcal{M}_1$ and $\mathcal{M}_2 = \{1,2,3,4\}$. It is easy to compute that  $R_{\text{DF}}(\mathcal{M}_1)=1.30826$ and $R_{\text{DF}}(\mathcal{M}_2) = 1.31576$.
\end{remark}

\section{DF with LDPC Codes} \label{sec:simulations}
In the previous section, we computed achievable rates of different strategies and routes based on an information theoretic approach. In this section, we compare the different strategies and routes in a line network using practical low density parity check (LDPC) codes \cite{gallager62}\cite{mackay99} with incremental redundancy. The aims of this section are:
\begin{enumerate}
\item To illustrate that DF on the multiple relay channel is implementable.
\item To demonstrate that DF performs better than MH under certain network topologies.
\item To show that routing backward (away from the destination) can be good in DF.
\item To demonstrate that the NNSA route performs better than other routes using DF.
\end{enumerate}

\begin{figure}[t]
\centering
\includegraphics[width=0.6\linewidth]{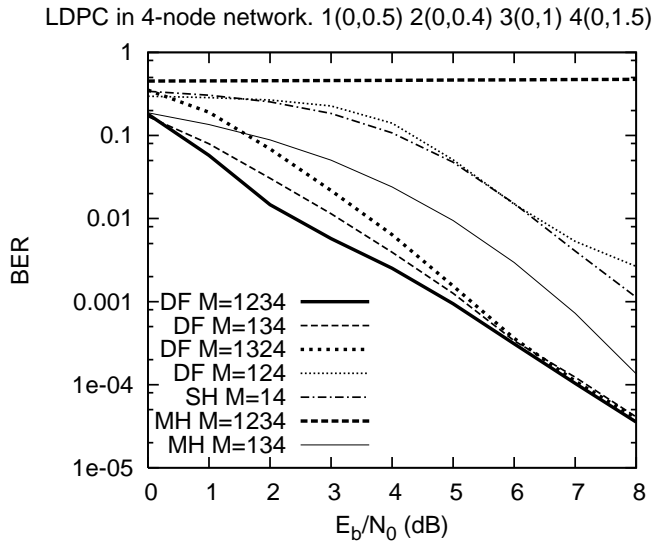}
\caption{Performance (information bit error rate (BER) versus transmit SNR) of different strategies on different routes in a 4-node network.}
\label{fig:mh-4-nodes-ldpc}
\end{figure}

From Fig.~\ref{fig:mh-4-nodes-ldpc}, we see that for a given route, DF performs better than MH. An interesting observation is that routing backward helps in DF but not MH. We find that the NNSA route (which is also the MSPA route), i.e., $\{1,2,3,4\}$, achieves the lowest BER compared to other routes using DF.

\begin{remark}
We note that the total transmit energy differs depending on the length of the route. One might argue that route $\{1,4\}$, though having a higher BER, is better as only $1/3$ power is consumed compared to route $\{1,2,3,4\}$. However, we stress that this paper finds a route that maximizes the transmission rate, given that each node must transmit within a given power constraint. Whether or not the node transmits, it is not important in the route comparison.
\end{remark}
\begin{remark}
We plotted BER versus SNR in Fig. ~\ref{fig:mh-4-nodes-ldpc}.  If the maximum raw channel data rate (in bps) is $\Psi_{max}$, 
then the throughput is $\Psi=(1-\text{PER})\cdot \Psi_{max}$, where PER is the packet error rate and depends on the BER and the packet size.
In simulations, we found that packet error rate (PER) had the same behavior as BER. 
%As erroneous packets can discarded at the application layer, BER might not be useful. 
%In this case,  $\text{throughput} = (1-\text{PER})$ gives the rate of correctly received packets at the destination.
\end{remark}

\begin{figure}[t]
\centering
\includegraphics[width=0.3\linewidth]{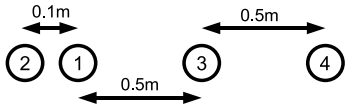}
\caption{Network topology.}
\label{fig:mh-4-nodes-ldpc-b}
\end{figure}

\section{Concluding Remarks}
We first showed that DF gives a significant transmission rate gain over MH, for an arbitrary route in the wireless network. We presented an algorithm, the nearest neighbor set algorithm (NNSA), to find optimal routes, which maximize the rates achievable by DF. As this algorithm, in the worst case, runs in factorial time, we designed a heuristic algorithm, the maximum sum-of-received-power algorithm (MSPA), that runs in polynomial time. We showed that the MSPA finds an optimal route when the nodes can only send independent codewords. However, unlike the NNA and the NNSA, the MSPA does not find an optimal route when the nodes are allowed to send arbitrarily correlated codewords. We implemented DF on practical networks using LDPC codes with incremental redundancy to compare different routes.

We would like to highlight that for a given route, the choice of coding strategies, be it MH or DF, does not affect the spatial re-use of the system. In both strategies, the nodes in the route (except the destination) transmit at the same power level. The difference lies in how the nodes decode the data.

We also note that there are some practical problems in implementing DF in large networks.  First, since real-world systems have finite transmit power constraints, nodes have a finite communication range, beyond which they cannot be heard. Second, large networks may be distributed over a wide area and cooperation over large distances may not be feasible.  The solution is to partition the network into clusters and perform cooperative coding locally in the cluster, e.g., \emph{local DF}, in which only nodes in a cluster cooperate with each other.

%\section*{Acknowledgment}
%The authors would like to thank Yap Kok Kiong for valuable comments and useful discussions.

\bibliography{bib}

\end{document}